\documentclass[aps,manuscript,overcite]{revtex}

\usepackage{graphics}

\input epsf

\tighten
\begin{document}
 


\title{From Big Crunch to Big Bang}

\author{ Justin Khoury$^1$, Burt A. Ovrut$^2$, Nathan Seiberg$^3$,
Paul J. Steinhardt$^1$ and Neil Turok$^{4}$}

\address{
$^1$ Joseph Henry Laboratories,
Princeton University,
Princeton, NJ 08544, USA \\
$^2$ Department of Physics, University of Pennsylvania,
Philadelphia, PA 19104-6396, USA\\
$^3$School of Natural Sciences, Institute for Advanced Study,
Princeton, NJ 08540\\
$^4$DAMTP, CMS, Wilberforce Road, Cambridge, CB3 0WA, UK
}

\maketitle

\begin{abstract}

We consider conditions under which a universe
contracting towards a big crunch  can
make a   transition to an expanding big bang universe.
A  promising example is 11-dimensional  M-theory
in which the eleventh dimension collapses, bounces, and
re-expands.
At the bounce, the model can reduce to a weakly coupled
heterotic string theory and, we conjecture, it 
may be possible to follow 
the transition from contraction to expansion.
The possibility opens the door to new classes of cosmological
models.  For example, we discuss how it suggests 
a major simplification and modification of the recently proposed 
ekpyrotic scenario.

\end{abstract} 
%

\section{Introduction}

Since the discovery of the cosmic microwave background, the 
predominant view has 
been that the universe originated from a cosmic singularity.
An important consequence is that  the universe has a finite 
age and a finite causal horizon distance.
For the standard hot big bang model, this leads to 
the horizon puzzle that inspired inflationary cosmology \cite{Gut}\null.  By 
introducing a period of superluminal expansion, inflation 
alleviates the horizon 
puzzle, but it is generally believed that an 
initial singularity is still required at the outset.

In this paper, we consider the possibility that the singularity is 
actually a  transition between a contracting big crunch
phase and an expanding big bang 
phase.  If true, the  universe may have existed for a semi-infinite  time
prior to the putative big bang.  The horizon puzzle would be nullified, 
eliminating one of the prime motivations for inflation.  The analysis
opens the door to alternative cosmologies with other solutions to 
the remaining cosmological puzzles.

The discussion in this paper
 focuses on  $d$-dimensional
 field theory and  is not specific to any particular cosmological
 model. A crucial role in our analysis is played by a massless scalar field
- a modulus. (The cosmology of such fields has been analyzed by many 
authors.\cite{Ven,moduli,banks})
We eschew any use of branes and strings until absolutely necessary.
String theory will become important at the point where the universe
bounces from contraction to expansion.

Here, our discussion is closely related to considerations of
the reversal problem in 
the pre-big bang scenarios of Veneziano, {\it et al.}\cite{Ven,pbbrev,Bru}
and related scenarios\cite{Kal,Eas,Lid,Luk}\null.
A notable difference, as we shall emphasize  below,
 is that the reversal in the  pre-big bang model
 occurs in the limit of
 strongly coupled string theory 
whereas we are interested in reversal in the limit of 
 weakly coupled string theory \cite{banks,kaloper}\null. 
 The differences have profound consequences for both cosmology and
fundamental physics.

In Section II, we show that reversal requires either  a violation of the 
null energy condition or passage through a singularity where the
scale factor shrinks to zero.  The remainder of the paper explores
the second possibility.
In Section III, we obtain the solutions for  a single 
scalar field evolving
in a contracting Friedmann-Robertson-Walker background
and use them to demonstrate the difference between our proposal
and the pre-big bang scenario.
In Section IV, we reformulate the theory in variables which 
are finite as the scale factor shrinks to zero and which suggest
a natural way to match solutions at the bounce.  Section V 
solves for the evolution before and after the bounce in cases
where the bounce is elastic and time-symmetric and cases where it
is inelastic and time-asymmetric.  Section VI compares the singularity
considered here to bounces and singularities considered in other contexts
in string theory.  Section VII focuses on the bounce itself where, 
we argue, string theory must play a critical role in passage through the 
singularity.  We conjecture that the cosmological
singularity  connects contracting
and expanding solutions in a manner analogous to the conifold and
flop transitions.
In Section VIII, we discuss the implications of our results 
for cosmology, particularly the ekpyrotic scenario \cite{ek1}\null. We suggest 
a major modification of the scenario in which the hot big bang 
phase is created in 
a collision between two boundary branes, 
removing the need for bulk branes.

\section{Reversal and the Null Energy Condition}

The reversal problem is notoriously difficult because
a violation of the null energy condition is required.
Consider a general 4d theory of scalar fields $\phi^K$
coupled to gravity.
We are  free to Weyl transform to the Einstein frame:
\begin{equation} \label{action}
S = \int d^4 x \sqrt{-g} \left\{ {1\over 16 \pi G_N} {\cal R} -
 \frac{1}{2}g^{\mu \nu} 
G_{IJ}(\phi^K) \partial_{\mu}\phi^I \partial_{\nu} \phi^J -V(\phi^K)
\right\},
\end{equation}
where $g_{\mu \nu}$ is the spacetime metric (in Minkowski space, 
$\eta_{\mu \nu} = (-1, \, +1, \, +1 \, +1)$), ${\cal R}$ is the 
Ricci scalar,
and $G_{IJ}(\phi^K)$ is the metric on field space.
We consider unitary theories in which $G_{IJ}$ is positive definite.
The energy momentum tensor is 
\begin{equation}
T_{\mu \nu} = G_{IJ} \partial_{\mu} \phi^I \partial_{\nu} \phi^J-
g_{\mu \nu} \left[{1\over 2} g^{\alpha \beta} G_{IJ} \partial_{\alpha}\phi^I
\partial_{\beta} \phi^J + V \right],
\end{equation}
where the repeated Greek or Roman indices follow the standard 
summation convention.
If we assume homogeneity and isotropy, then  $g_{\mu \nu}$ is 
a Friedmann-Robertson-Walker metric and 
the energy density $\rho$ and the pressure $p$ are
\begin{eqnarray}
\rho & = & T_{00} =  \frac{1}{2} G_{IJ} \dot{\phi}^I \dot{\phi}^J + V \\
p &=& -{1\over 3}g^{ij} T_{ij} = \frac{1}{2} G_{IJ} \dot{\phi}^I \dot{\phi}^J - V, 
\end{eqnarray}
where dots denote proper time derivatives, and $i,j=1,2,3$.

A problem arises because scalar fields satisfy the null energy condition:
\begin{equation}
\rho + p = G_{IJ} \dot{\phi}^I \dot{\phi}^J \ge 0.
\end{equation}
In a Friedmann-Robertson-Walker universe, the expansion rate is 
set by the parameter $H= \dot{a}/a$, where $a$ is the scale factor.
The time variation of $H$ assuming a flat  universe is given by 
\begin{equation}
\dot{H} = - 4 \pi G_N (\rho+p)  =  -4 \pi G_N G_{IJ} \dot{\phi}^I \dot{\phi}^J 
\leq  0.
\end{equation}
If $\dot{H} \leq 0$, 
then reversal from contraction ($H<0$) to expansion ($H>0$) is not possible.
Hence, we obtain the theorem:
Given a flat universe and a unitary theory with terms second order
in field derivatives, then 
the contracting big crunch phase and the expanding big bang phase
are separated by a singularity.
Similar theorems and considerations of ways of circumventing them 
have been discussed in the literature \cite{Bru,Kal,Eas,Bru2}\null.

\section{Two approaches to the reversal problem  }

Both the pre-big bang model and the scenario we consider entail
the problem of reversal from contraction to expansion in the 
Einstein frame.
In pre-big bang models, the common approach has been to consider 
violating one of the assumptions of the theorem derived
above \cite{pbbrev,Bru,Kal,Eas,Bru2}\null. For example,
by introducing higher derivative terms in the action,
the null energy condition can be violated and reversal  might 
take place
without reaching the singularity.  In this paper, we consider the 
alternative possibility of proceeding to $a=0$ and bouncing 
without introducing new terms in the action.

Both approaches can be modeled by an action with a scalar field
$\phi$ plus gravity:
\begin{equation} \label{single}
S = \int d^d x \sqrt{-g} \left\{ {\cal R}(g) -
 \frac{1}{2}g^{\mu \nu} 
\partial_{\mu}\phi \partial_{\nu} \phi -V(\phi)
\right\},
\end{equation}
where ${\cal R}(g)$ is the Ricci scalar based on the metric $g_{\mu \nu}$.
Here we have  generalized to 
$d$-dimensions and simplified to the case
of a single field $\phi$, and we have chosen units in which
the coefficient of ${\cal R}$ in the $d-$dimensional Einstein-frame
Lagrangian is unity.
For the moment, we consider the case where
the potential $V(\phi)=0$.
This kind of model arises in the low energy approximation to 
Type IIA and IIB string theory for $d=10$.
It also arises whenever a $d+1$-dimensional gravity theory is 
compactified to $d$-dimensions, where $\phi$ (the `radion') 
determines the length
of the
compact dimension.

We further simplify by restricting ourselves to a mini-superspace 
consisting of spatially flat,
homogeneous and isotropic solutions with metric
$ds^2 = a^2(t) \left[-N^2(t) dt^2 
+ \sum_{i=1}^{d-1} (d x^i)^2 \right]$.
In the gauge $N=1$, $t$ is conformal time.
Then, the solutions to the equations of motion  up to a shift
in $t$ are:
\begin{equation} \label{sol}
a= a(1) |t|^{\frac{1}{d-2}} \; \;  {\rm and} \; \; \phi= \phi(1) + \eta
\sqrt{\frac{2(d-1)}{d-2}} \, {\rm log}\, |t|,
\end{equation}
where $\eta = \pm 1$ and $a(1)$ and $\phi(1)$ are integration constants
set by the initial conditions.
Each solution has two branches.  For $t<0$, the Universe contracts to 
a big crunch as $t\rightarrow 0^-$. For $t>0$, the Universe expands from 
a big bang beginning at $t \rightarrow  0^+$.  

At $t=0$, the solutions are singular and $\phi \rightarrow \mp 
\infty$. In the case of Type IIA (or heterotic) string theory
in $d=10$, the string coupling is $g_s = e^\phi$
(see discussion of Eq.~(\ref{eq:stringmet}) below),
so the two solutions at the bounce
(as $ t \rightarrow 0$)  correspond to 
weak and strong coupling, respectively.

We will find it  useful to re-express the model 
in terms of the ``string metric" $g^{(s)}_{\mu\nu}=e^{\phi/c} g_{\mu\nu}$
where $c=\sqrt{(d-2)/2}$.
The action based on the string metric is:
\begin{equation}
 \label{eq:stringactd}
\int d^{d}x \sqrt{- g^{(s)}}e^{-c \phi}\left( R (g^{(s)})+ c^2
 g^{(s)\mu\nu} \partial_\mu \phi \partial_\nu \phi \right).
\end{equation}
This  definition 
of the string metric agrees with the standard metric for $d=10$ and the 
string metric for $d=4$ as defined in the pre-big bang 
literature \cite{Ven,pbbrev}\null.

The two solutions in Eq.~(\ref{sol})  are easily transformed to the string frame
$g^{(s)}_{\mu\nu}=a_s^2\eta_{\mu\nu}$, where we 
can compare them to solutions in 
the pre-bang scenario.  
We can express the solutions in terms of string-frame FRW
time $\tau_s$, where  $d\tau_s=a_s dt$:
\begin{eqnarray} \label{pbbans}
  a_s & = &a_s(1)|\tau_s|^{\eta \over \sqrt{d-1}} \nonumber \\
   \phi &= &\phi(0)+\sqrt{2\over d-2}(\eta\sqrt{d-1}-1) \log |\tau_s|.
   \end{eqnarray}
Using these we find
\begin{eqnarray}
     H_s & \equiv & {\dot{a}_s\over a_s}={\eta \over \sqrt{d-1}} {1 \over \tau_s}  \nonumber\cr
     \dot{\phi} & = &\sqrt{2\over d-2}(\eta\sqrt{d-1}-1){1\over \tau_s } \nonumber \cr
       \dot{\bar \phi} & \equiv & \dot{\phi} - (d-1)\sqrt{2\over d-2} H_s = - \sqrt{2\over
	d-2} {1\over \tau_s}, \cr 
\end{eqnarray}
where dot denotes differentiation with respect to $\tau_s$.

There is a key difference between the pre-big bang scenario and the 
reversal considered here.
We approach weak coupling as 
$ t \rightarrow 0^-$ corresponding to the $\eta=+1$  branch of
Eq.~(\ref{sol}).  The pre-big bang scenario approaches strong coupling
as $t \rightarrow 0^-$ corresponding to the $\eta=-1$ solution.
The two scenarios may be compared
by  mapping their trajectories in the plane spanned by $\dot{\bar{\phi}}$ and $ 
 H_s$,
a method  commonly introduced\cite{pbbrev}
to describe the pre-big bang picture in $d=4$.
Note that the ratio 
\begin{equation}
\frac{ \dot{\bar{\phi}}}{  H_s } = - \eta \, \sqrt{\frac{2 (d-1)}{d-2}}
\end{equation} 
 is negative for $d>2$ if $\eta = +1$ (our
solution) and positive if  $\eta =-1$.  This expression 
describes the four solutions shown in Fig.~1.
As noted above, for the pre-big bang model, 
the $\eta=-1$ solution in which $\phi$ runs to $+ \infty$ (strong 
coupling) is chosen for $t<0$.
It has been proposed \cite{Bru,Kal,Eas,Bru2}
that new terms in the action 
 appear in the strong coupling limit that violate the 
null energy condition
 ({\it e.g.}, by introducing high derivative
interactions and potentials) making it possible to avoid $\phi$
running off to $+ \infty$, as shown by the dashed curve.  
Whatever physics is involved, it is presumed to freeze the dilaton (so
Eq.~(\ref{sol}) is no longer applicable) and create radiation  that 
dominates the Universe.
Nevertheless, $H_s$ is positive 
and $\dot{\bar{\phi}} = - (d-1) \sqrt{2/(d-2)}
H_s$ (assuming $\phi$ is frozen) is negative.
Hence, the Universe joins onto a path similar to what is shown in the Figure.
By contrast, the trajectory proposed in this paper
maintains $\eta  = +1$ throughout. This is a fundamental difference
that distinguishes everything we say in the remainder of this paper
from the pre-big bang scenario.

\begin{figure}
\begin{center}
{\par\centering \resizebox*{6.5in}{3.1in}{\includegraphics{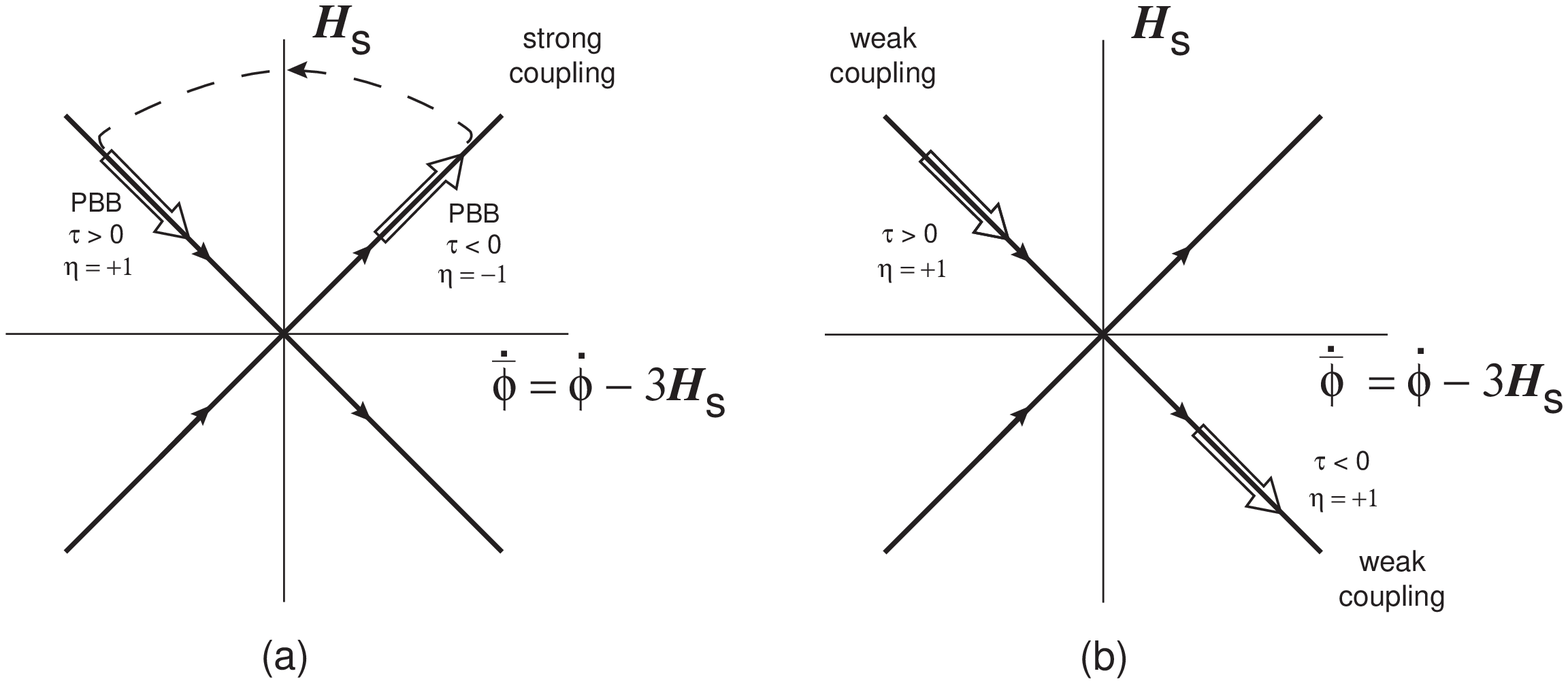}} \par}
\end{center}
\caption{
Phase diagrams for $d=4$ comparing the (a) pre-big bang (PBB) model with
(b) the bounce scenario considered here.  
The four rays connected at the origin represent the four solutions 
to the potential-less equations of motion.  
The large arrows indicate the two solutions that are joined together 
in each of the two cosmologies.  
Reversal from contraction to expansion connects the two weak coupling
regimes in (b).
}
\end{figure}

\section{How singular is the singularity?}

A key step in tracking the  universe across the bounce at $a=0$
is to find
variables which are finite as $t \rightarrow 0$.
Consider the change of variables \cite{ek1,kirklin}\null:
\begin{eqnarray}
a_0 & = &  a^{\frac{d-2}{2}}\left( {\rm e}^{-\gamma \phi} + {\rm e}^{\gamma \phi}
\right), \nonumber
\\
a_1 & = &  a^{\frac{d-2}{2}}\left( {\rm e}^{-\gamma \phi} - {\rm e}^{\gamma \phi}
\right), \nonumber
\\
a_{\pm} & = &  \frac{1}{2} ( a_0 \pm a_1) = 
a^{d-2\over 2}e^{\mp\gamma\phi} 
\label{defns}
\end{eqnarray}
where $\gamma = \sqrt{(d-2)/8(d-1)}$.
 Their range 
for $a>0$ and $\phi$ real is the quadrant
$a_\pm >0$ or $a_0\ge |a_1|$. 

The effective Lagrangian for $V=0$  is transformed to \cite{ek1}\null:
\begin{equation} \label{eq:apm}
{d-1\over N(d-2)}[-{a_0'}^2+ {a_1'}^2]
=
-\frac{4 (d-1)}{N(d-2)} {a}_+' {a}_-',
\end{equation}
where primes denote derivatives with respect to conformal time $t$.
In the moduli space spanned by $(a_0,a_1)$ we identify $a_0$ ($a_1$) as a
time-like (space-like)
variable and $a_{\pm}$ as light-cone coordinates.
We shall consider 
trajectories which bounce at 
$a =0$ corresponding to
a point on the moduli
space boundary
$a_0=a_1 \ne 0$. Without loss of generality the value of 
$t$ at the bounce can be chosen to be
$t=0$.

We cannot describe exactly what occurs at $t=0$. 
However, what is 
encouraging is that we have found  a choice of variables, 
$a_{0,1}$ that 
remain finite for $t<0$ and $t>0$, and there appears 
to be a  natural  way to match at $t=0$.
It is instructive to change  variables for the solution
to Eq.~(\ref{sol}) with $\eta=+1$  to
\begin{eqnarray} \label{barg}
\psi & = & {\rm e}^{\gamma \phi}  \nonumber \\
\bar{g}_{\mu \nu} & = & \psi^{-4/(d-2)} g_{\mu \nu}.
\end{eqnarray}
This 
leads to a reformulated action
\begin{equation} \label{eq1}
S = \int d^d x \sqrt{-\bar{g}} \, \psi^2 {\cal R}(\bar{g}),
\end{equation}
with no kinetic term for $\psi$.
We  use a parameterization such that the coefficient of 
${\cal R}$ in Eq.~(\ref{eq1}) is  $\psi^2$ to ensure that it is always 
positive.

The scale factor of the metric $\bar{g}_{\mu \nu}$ is $\bar{a} =
\psi^{-2/(d-2)} a$.   The solution to the equations of motion in 
Eq.~(\ref{sol}) (with $\eta=+1$) become 
\begin{equation} \label{eq:barsol}
\bar{a} = A \; \; {\rm and} \; \; \psi= B|t|^{\frac{1}2{}}
\end{equation}
with $A$ and $B$ positive constants. 
By rescaling the $d$ dimensional coordinates one can always set $A=1$.
In terms of the original metric
$g_{\mu \nu}$, the Universe shrinks to a point at $t=0$.
However, we see that the metric $\bar{g}_{\mu \nu}$ is smooth there.

One should not read too much into this.  The change of variables
does not make the problem entirely regular.  First and foremost, 
since
$\psi(t=0) =0$, the Planck scale vanishes at the bounce
in these coordinates. Hence, there is the concern that
quantum fluctuations become uncontrolled at $t=0$.  
Note also that ${\psi'}$ is singular at $t=0$, so higher 
dimension operators are also important at the bounce.
Therefore, even in these variables, we 
see the importance of going 
beyond the field theoretic descriptions to understand
the physics at $t=0$.

We must assume that the field theories considered here are low
energy approximations to some more fundamental theory. The action
in Eq.~(\ref{eq1})  in terms of $\psi$ and $\bar{g}_{\mu \nu}$ is
just the Einstein-Hilbert action for $(d+1)$-dimensional gravity
theory compactified to $d$ dimensions. To see that, consider the
$d+1$-dimensional metric
\begin{equation}  \label{met}
ds^2 = \psi(x)^4 dw^2 + \bar{g}_{\mu \nu} d x^{\mu} d x^{\nu}
\end{equation}
and let it depend only on the $d$-dimensional variables $x^{\mu}$.
(For simplicity, we neglect the vector field arising from the
$\mu-w$ components of the metric.) A straightforward calculation
leads to
\begin{equation} \label{tran}
\sqrt{-g^{(d+1)}} {\cal R}(g^{(d+1)}) = \psi^2 \sqrt{-\bar{g}}
{\cal R}(\bar{g}) - 2 \partial_{\mu}( \sqrt{-\bar{g}}
\bar{g}^{\mu \nu} \partial_{\nu} \psi^2),
\end{equation}
where $g^{(d+1)}$ is the metric in $d+1$-dimensions.  Constraining
$w$ to lie in the range $[0,\, 1]$, the $d+1$-dimensional
Hilbert-Einstein action is reduced to Eq.~(\ref{eq1}). For
example, the compactification can be on a circle, as in
Kaluza-Klein theory,  or on an interval, where  $\phi$ or $\psi$
can be interpreted as a radion. If the compactified dimension is
a line segment there are two boundary branes at the ends
\cite{REFBRANE,HW,polch}\null. Then, $a=0$ corresponds to a
circle collapsing to a point or the branes colliding at $t=0$.

\begin{figure}
\begin{center}
{\par\centering \resizebox*{4in}{4in}{\includegraphics{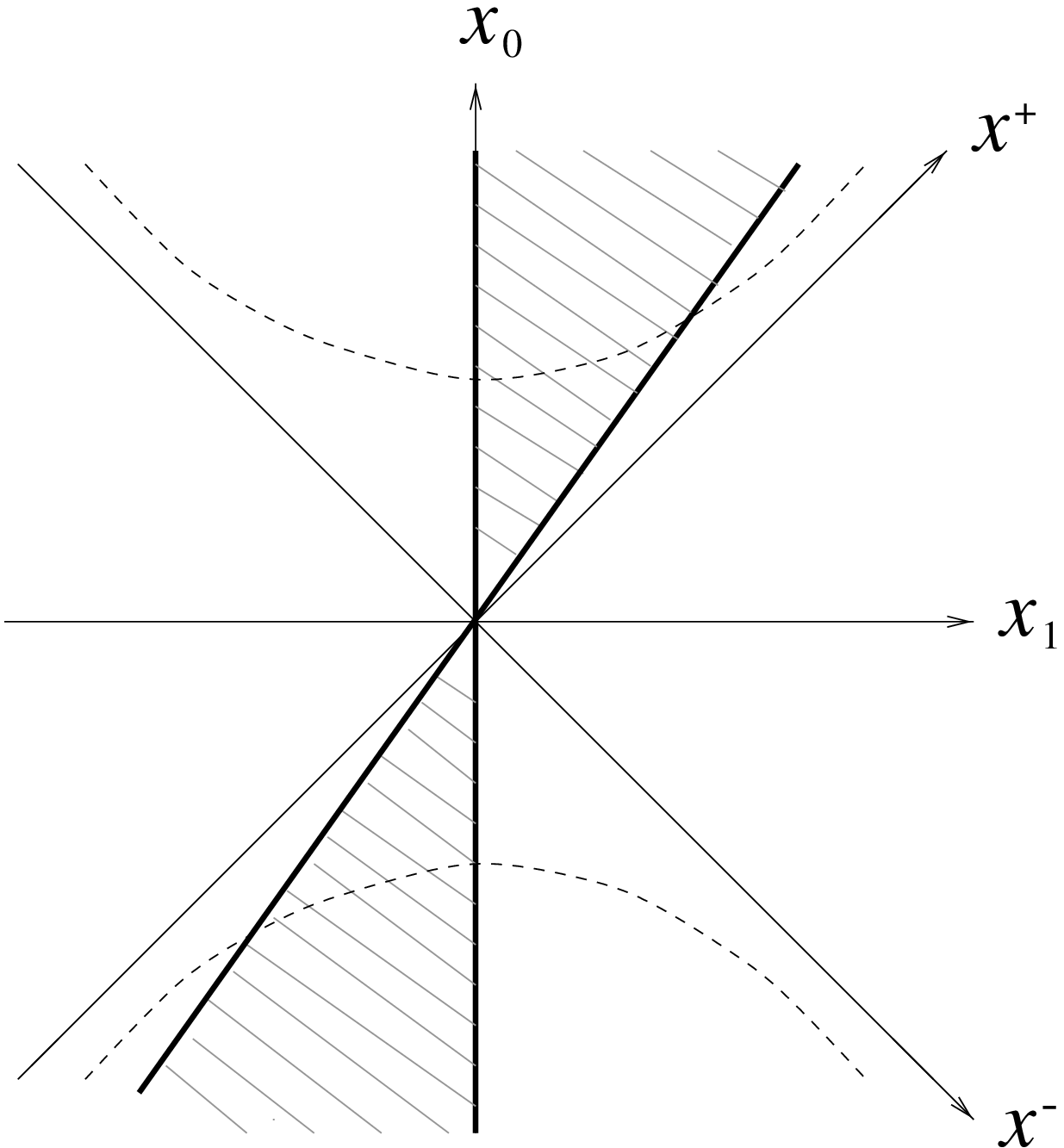}} \par}
\end{center}
\caption{
Sketch of the compactified Milne universe (hatched region)
embedded in a Minkowski 
background, where $x^0$ and $x^1$ are the time and space coordinates.
The dashed surfaces are surfaces of constant $t$.}
\label{milne}
\end{figure}

Substituting Eq.~(\ref{eq:barsol}) into the metric, we obtain
\begin{equation}
ds^2 = B^4 t^2 dw^2 + \eta_{\mu \nu} dx^{\mu} dx^{\nu}.
\end{equation}
The space-time is remarkably simple.
It is simply ${\bf R}^{d-1} \times {\cal M}^2$,
where the $d-1$ dimensions are Euclidean and 
${\cal M}^2$ is a  $2-d$ compactified  Milne universe (Figure 
\ref{milne}) with $ds^2 =-dt^2 
+B^4t^2 dw^2$.
Each branch of our solutions spans a wedge in Minkowski 
space
compactified 
on an interval $w\in [0,\, 1]$  with endpoints identified. 
Equivalently, 
if the metric 
is re-expressed in Minkowski lightcone coordinates
$d s^2 = d x^+ d x^-$
where $x^\pm= \pm te^{\pm B^2 w}$, then  ${\cal M}^2$
corresponds to flat Minkowski space modded out by
the boost $x^+ \to  {\rm exp}(B^2)\,  x^+$, $x^- \to
{\rm exp}(-{B^2})\, x^-$.
Such a compactified Milne universe has been discussed 
G. Horowitz and A. Steif.\cite{Horo}
Our bounce connects two 
branches of  ${\cal M}^2$  at $t=0$. As mentioned, if the extra dimension
is a circle, it contracts to zero at $t=0$, and re-expands. If the extra
dimension is an interval, the two boundary branes follow the 
heavy lines in Figure \ref{milne}, bouncing off each other.
Equivalently, as the Figure suggests, one can say that the
two boundary branes meet and pass through one another.
The variables $a_0$ and $a_1$ defined 
above are given by $a_0=1+B^2 |t|$, 
$a_1=1-B^2 |t|$, so these bounce at $t=0$. 
Finally, note that the proper distance between the branes in the
Milne metric is $B^2 |t|$, and we see the physical interpretation 
of the constant $B^2$ as the magnitude of the relative
velocity of the branes. 

In the usual Kaluza-Klein reduction from $d+1$ to $d$ dimensions
the variables defined in (\ref{defns}) parameterize the geometry as follows.
The scale factor of the
noncompact $d$ dimensional space as
measured by the canonical $d+1$ dimensional metric 
is $\bar a= a_+^{2/(d-2)}$, which has been set to unity. The 
size of the extra dimension is proportional to
$a_-/a_+$, which can take on any positive value. Thus the 
range of $\phi$ is $-\infty <\phi<\infty$.
Other  $d+1$ dimensional theories can
reduce to the same $d$ dimensional effective field theory,
but the geometrical meaning of the $a$, $\phi$ variables
and their range may differ. For example, consider 
$AdS_5$ bounded by a positive and a negative tension brane.
The induced scale factor on  the positive (negative)
tension brane is $a_0$ ($a_1$), with
$0<a_1<a_0$, so that $\phi$ is restricted to be less than zero. The
distance between the 
branes is proportional to log$(a_0/a_1)$, which 
agrees with the  Kaluza-Klein 
result at short distances 
where the variation of the warp factor is
negligible.
We note that more general compactifications with additional 
dimensions lead to 
more complicated actions which depend on several moduli.
If the moduli space can be reformulated in terms of variables 
analogous to $a_{\pm}$ that   are finite at the bounce, 
a similar analysis should hold.  
Alternatively, the bounce trajectories are restricted to cases
where the time-derivatives of the additional fields are zero and
 the theory reduces to the current examples.
However, the simple interpretation of
$\bar{g}_{\mu \nu}$ in Eq.~(\ref{met}) as a time-independent metric
is only valid for compactifications of a single dimension.

When the theory in Eq.~(\ref{eq1})
is derived from compactification as in
Eq.~(\ref{tran}),  the bounce
solution corresponds to shrinking the compact  dimension to zero size
and then expanding it again.  
(In the work of Brandenberger, Vafa, and Tseytlin \cite{Bra,Tse}\null,
they considered the situation where one spatial dimension 
collapsed and a different one opened up. This is also
an interesting possibility which may be equally good for our purposes.)
Throughout this
process the metric in the noncompact dimensions as measured by
the $d+1$-dimensional metric in Eq.~(\ref{met}) is
unchanged.  Such an
intuitive picture suggests that indeed the two branches of the
solution in Eq.~(\ref{sol})
(or in terms of the coordinates $\bar a$,
$\psi$ in Eq.~(\ref{eq1})) are indeed connected.
However, 
it should be stressed that the dimensional reduction from $d+1$
dimensions to $d$ dimensions makes the bounce natural but it does
not prove that it exists.  

\section{Approaching the bounce}

We have the machinery in hand to track the evolution as the 
Universe approaches the bounce (or rebounds afterwards).
From varying the action in 
   Eq.~(\ref{eq:apm}) with respect to $N$, we
obtain 
the constraint, $-{{a}_0'}^2 + {a_1'}^2 =0$.  
(Expressed in terms of $a$ and $\phi$, this corresponds to the 
Friedmann equation.)
Consequently, we are only permitted solutions where  
${a}_0' = \pm {a}_1'$. The minus sign solution must apply
if the branes are to collide. The incoming trajectory 
intersects the light-like 
boundary of moduli space $a_0 = a_1$ along a light-like
trajectory.
If we assume that no radiation is 
produced, then
to satisfy the energy 
constraint, the solution after collision must also be 
light-like. There then appears only one natural possibility for 
the trajectory to follow, which is to bounce straight back off the
light cone,  
${a}_0' \leftrightarrow {a}_1'$, as occurs in 
the Milne universe example explained above.

Returning to the Lagrangian in Eq.~(\ref{eq:apm}),  
we now add a potential term,
$-Na^dV(\phi)=-N(a_+a_-)^{d\over d-2}V\left({1 \over 2\gamma} \log
(a_-/a_+)\right)$.  Up to an unimportant constant,
the total Lagrangian 
becomes:
\begin{equation} \label{efflagp}
 - {1\over N} a_+'  a_-' -N(a_+a_-)^{d\over
  d-2}F(a_-/a_+)
\end{equation}
  where the function $F$ is related to the potential.
Since
our convention  is  that the weak coupling region is $\phi
\rightarrow -\infty$,   the potential should vanish in that limit or,  
equivalently,
$F(a_-/a_+)$ should vanish for small values of its argument.

As an exercise, it is instructive to consider a case where
the equations of motion are exactly solvable:
\begin{equation} \label{spef}
 F(a_-/a_+)=\epsilon \left({a_-\over a_+}\right)^{d\over
 d-2}
\end{equation}
where $\epsilon= \pm 1$.  This example
 corresponds to $V \sim \epsilon
\exp\left({d\over\sqrt{2(d-1)(d-2)}}\phi\right)$.  In the gauge
$N=1$ the solution up to a shift of $t$ around $t=0$ is
\begin{eqnarray} \label{soleomg}
 a_- &  = & p|t| \cr
 a_+ & = & a_+(0) +\epsilon {d-2\over 3d-2} p^{d+2\over d-2} |t|^{3d-2\over
 d-2},
 \end{eqnarray}
 where $p$ is an arbitrary positive constant.

Consider first the case of $\epsilon$ positive, i.e.\ a positive
potential.
The Universe has  $a=0$ at $t=0$ where $a_-=0$.
For positive $t$, the universe
expands to infinite size.  $\phi$ on
the other hand  approaches  $-\infty$
at $t \rightarrow 0^+$,  it increases, reaches a maximum value at some
fixed $t$, and then slides back to $-\infty$ as $t\to \infty$. For
negative times the picture is symmetric.  In particular, the 
universe is contracting as $t \rightarrow 0^-$. 

\begin{figure}
\begin{center}
{\par\centering {\includegraphics{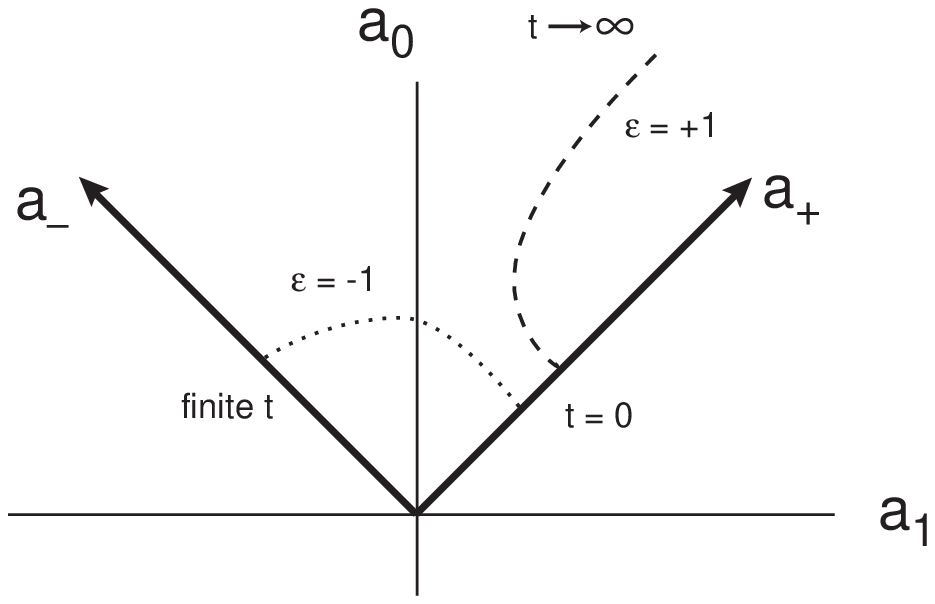}} \par}
\end{center}
\caption{
The moduli space in $a_0$ and $a_1$ or, equivalently, light-cone
coordinates $a_{\pm}$.  The physical regime is the upper light-cone
(quadrant).  The two trajectories 
correspond to the exact solutions  for the potential discussed in
the text for $t >0$. 
The bounce occurs at $t=0$.
By construction the solutions are time-symmetric.
The dashed solution corresponds to $\epsilon =+1$  and the
dotted corresponds to $\epsilon=-1$. 
}
\label{figure3}
\end{figure}

For negative $\epsilon$ the potential is negative.
In this case
the solution  Eq.~(\ref{soleomg})  cannot be trusted beyond a critical value
of $t=t_0$ where 
$a_+$ vanishes. 
At that point $a=0$ and
$\phi=+\infty$.  Since the potential is not bounded from below,
it is not surprising that $\phi$ reaches $\infty$ in finite
time.  In the brane picture, the repulsion makes the higher
dimensional space infinite in a finite time.  We have no reason
to expect another bounce at this point.  Again, the picture is
symmetric around $t=0$. The two solutions are represented in 
Figure \ref{figure3}.

In the examples considered thus far, the bounce at $\phi
\rightarrow -\infty$ is time-symmetric. The potential is taken to
vanish in that limit, and the trajectory in the $(a_0,a_1)$-plane
intersects the boundary of moduli space $a_0=a_1$ along a
light-like direction. After the bounce it simply reverses,
corresponding to the matching condition ${a}_{0,1}'({\rm out}) =
-{a}_{0,1}'({\rm in})$. This could be described as an {\it
elastic} collision, since the internal states of the two branes
are unchanged after collision.

As the velocity approaches zero, the boundary brane collision may
be nearly elastic, resulting in  no radiation being  produced on
the branes. But at finite velocity, we should expect entropy
production as radiation modes are excited both in the bulk and on
the colliding branes.

Let us consider the description of fluids produced on
the branes at the collision.
The action for a fluid in a background metric $\hat{g}_{\mu \nu}$
is $-\int d^d x \sqrt{-\hat{g}} \rho$, where $\rho$ is implicitly
determined in terms of $\hat{g}_{\mu \nu}$ by the fluid equations. 
In the present context, where the matter couples to the higher 
dimensional metric, we should take  $\hat{g}_{\mu \nu}$ to be 
$\bar{g}_{\mu \nu}$ given in Eq.~(\ref{barg}), rather than the
Einstein-frame four dimensional metric $g_{\mu \nu}$. 
This difference is very
important. Whereas the Einstein frame scale factor $a$ 
vanishes at the singularity, the scale factor $\bar{a}$ is finite there.
In consequence, fluids coupling to $\bar{a}$ have finite density
and temperature at the singularity. The
usual infinite blue-shift caused by the vanishing of $a$ is
precisely cancelled by an infinite `fifth force' red-shift
due to the coupling to $\phi$, as
$\phi \rightarrow -\infty$. We see once more that within the context
we are discussing, the big crunch/big bang singularity is 
remarkably non-singular. 

Let us consider as an example
the case where the incoming state has no
radiation, and the potential $V(\phi)$ asymptotes to zero as
$\phi \to -\infty$, and remains zero after the collision.
Assume also that a small amount of
radiation is produced, so the collision is
slightly `inelastic'. The `elastic' matching condition discussed
above, ${a}_{0,1}'({\rm out}) = -{a}_{0,1}'({\rm in})$, cannot
apply, since it would be incompatible with 
the Friedmann
constraint, which reads
\begin{equation}
{a}_0'({\rm out})^2 -{a}_1'({\rm out})^2 =  \left({d-2\over d-1} \right)
\rho(\bar{a}) \bar{a}^d. 
\label{eq:fc}
\end{equation}
As stressed above, each term in this equation is perfectly
finite at $a=0$ (the `singularity'). But the presence of
the positive radiation/matter density term on the right hand side 
means that 
the outgoing
trajectory must be time-like in the $(a_{0},a_1)$-plane. 

The details of the microscopic physics determine the amount of
radiation which is generated by the collision.  In terms of
the long distance effective theory that we have been using the
microscopic physics also determines the precise boundary
conditions on $a_0'$ and $a_1'$.  If before the collision the
system has no radiation and the potential vanishes for $\phi \to
-\infty$, the trajectory in field space hits the boundary along a
light-like curve. As we said, because of Eq. (\ref{eq:fc}) if
radiation is being generated, it bounces off the boundary along a
time-like curve and the trajectory is not time-symmetric.

This discussion will be further elaborated upon in
Ref.~(\ref{lyth})\null.

\section{Distance to the Singularity}

When referring to the moduli space  in string theory, one
usually has in mind the moduli of the compact dimensions,
keeping the noncompact dimensions unchanged. In particular, the
scale size of the noncompact dimensions $a$ is not usually considered
to be one of the
coordinates on moduli space.  Most of the singularities which are
studied in string theory are at finite distance in moduli space. At
such a singularity the presence of gravity can be neglected, and
the essential physics of the singularity is described by local
quantum field theory.  The latter can be either a weakly coupled
quantum field theory with new light degrees of freedom which
become massless at the singularity, or a strongly coupled quantum
field theory at a nontrivial fixed point of the renormalization
group.  A typical example of such a singularity is the small
$E_8$ instanton transition in which a bulk brane hits the 
boundary brane \cite{ganor,SW,pantev}\null.
 This is the singularity which was proposed to be
the 
initiation of the big bang phase in the
ekpyrotic model \cite{ek1}\null.

The singularity of interest here is of a totally different
nature.  We are interested in the singularity at $\phi=-\infty$.
The metric on $(a,\phi)$ space is given from the kinetic terms
in the action (\ref{eq:apm}). After a trivial scaling of $a$ the line
element is 
proportional to
$-da^2+a^2 d\phi^2$.
For fixed $a$, 
the singularity is clearly at infinite distance in
moduli space.  The way we manage to reach it at finite time is to
consider a motion not only in the $\phi$ moduli space
but in an extended space
including also the scale size of the noncompact dimensions $a$.
Since this extended space has Lorentzian signature, the proper distance
to $(\phi=-\infty,a=0)$ can be finite even when it is infinite to
any generic $a$.

The fact that $a$ vanishes at the singularity has profound
implications.  Unlike the other singularities which are field
theoretic, here gravity cannot be ignored.  Therefore, the
physics of the singularity cannot be described by local quantum
field theory coupled to weakly coupled gravity.  It is an
important challenge to find other such singularities and to
describe them in detail.

\section{Passage through the Singularity and the Role of String Theory}

To prove that the transition past  $a=0$ can occur smoothly,
 one must have a consistent
theory at short distances  and complete
control of the dynamics at the singularity.
Here is where string theory becomes an essential element.
To determine what happens at $a=0$,
it is  natural to try to embed
our solution in string theory which provides a complete theory of
quantum gravity.

Our equations can be embedded in string theory in several
different ways.  The most straightforward way is to embed it in
type IIA or the heterotic string in $d=10$ by identifying $\phi$
with the dilaton.  As  pointed out in the discussion following
Eq.~(\ref{met}), $\bar g_{\mu\nu}$ of Eq.~(\ref{barg}) is
the ten-dimensional metric measured in M theory units. So our
background is M theory on ${\bf R}^9\times {\cal  M}^2$, 
where ${\cal M}^2$ is the 2d compactified Milne space described
by the metric in Eq.~(\ref{met}).

Is our background a solution of the M theory equations of motion?
The fact that the M theory metric $(\psi^4,\bar g_{\mu\nu})$ is
flat might suggest that the answer to this question is positive.
However, since the background is obtained by modding flat
eleven-dimensional space by a boost $x^\pm \sim e^{\pm B^2} x^\pm$,
we should be more careful.  Spin half fields transform under this
operation as $F_\pm\to \zeta e^{\mp {1\over 2} B^2} F_\pm$ where
$\zeta=\pm 1$ is a choice of spin structure.  Therefore, there is
no covariantly constant spinor, our background breaks
supersymmetry, and it is not clear whether the quantum equations
of motion are satisfied.  For $|t|\to \infty$, where the
circumference of the circle is large this breaking is small and we
have a good approximation to a solution of the equations of
motion.  For small $|t|$ near the singularity the quantum effects
become large and a more careful analysis is needed.

Attempting to proceed to small $t$,  it is natural to change
variables to the string metric $g_{\mu\nu}^{(s)}=\psi^2 \bar
g_{\mu\nu}=\psi^{3\over 2} g_{\mu\nu}$.   Let $\phi={3\over
2}\log \psi^2$ be the dilaton in terms of which the action is
\begin{equation} \label{eq:stringact}
\int d^{10}x \sqrt{- g^{(s)}}e^{-2\phi}\left( R (g^{(s)})+4
 g^{(s)\mu\nu} \partial_\mu \phi \partial_\nu \phi \right).
 \end{equation}
The solution of the equations of motion is $\psi
\sim |t|^{1\over 2}$, and using the relations between the various
metrics
\begin{eqnarray} \label{eq:stringmet}
 g_{\mu\nu}^{(s)}& =  \psi^2 \bar g_{\mu\nu} =\alpha
 |t|\eta_{\mu\nu} \; \; {\rm and}
 \cr
 \nonumber  \\
 g_s^2& = e^{2\phi}=\psi^6 =\beta |t|^3
\end{eqnarray}
where  $g_s$ is the string coupling and $\alpha$ and $\beta$ are
arbitrary positive constants.  In terms of $\tau_s\sim t^{3\over
2}$, the string metric is $ds^2_s\sim -d\tau_s^2+ \tau_s^{2\over
3}\sum_{i=1}^9 (dx^i)^2$ and $g_s\sim |\tau_s|$. Note that the
string coupling vanishes at the singularity at $\tau_s =0$. It is
easy to see that the results in Eq.~(\ref{eq:stringmet})
satisfy the string equations of motion to leading order in
$\alpha'$ and $g_s$ (where we use the string metric
$g_{\mu\nu}^{(s)}$)
 \begin{eqnarray} 
R_{\mu\nu}(g_{\mu\nu}^{(s)})& = &-2\nabla_\mu
\nabla_\nu \phi
  ={3\over 2 t^2}
 (\eta_{\mu\nu} + 4\delta_{\mu 0} \delta_{\nu 0} )
 \nonumber
 \\
 \nabla_\mu \phi \nabla^\mu \phi & = &  {1\over 2} \nabla^2 \phi =
 -{9\over 4 |t|^3}
 \end{eqnarray}
and, therefore, lead to a conformal field theory to leading order
in $\alpha'$.  By choosing the constants $\alpha$ and $\beta$
appropriately, we can make the range
of validity
of this
approximation arbitrarily large although not
to $t=0$.

In  the long time limit $|t|\to \infty$, in the string frame
the Universe expands and becomes large. The string coupling 
$g_s$ also becomes large. However, 
the theory is still manageable.  In Type IIA theory, the
theory becomes M-theory in eleven dimensions where the size of the
eleventh dimension is large \cite{polch}\null.

We can also consider type IIB theory on our background.  The low
energy theory is still of the form in Eq.~(\ref{eq:stringact}), and its
solution can be expressed using the string theory variables as in
Eq.~(\ref{eq:stringmet}).
However, we no longer have the argument for the
bounce which is based on the compactification of a higher
dimensional theory. Still we can examine the behavior of each
branch of the solution.  At long time the string coupling is
large and we can use S-duality to transform the solution to
another weakly coupled description with $g_s^2 = {1\over
\beta |t|^3}$. The canonical Einstein metric,
$g_{\mu\nu}=\alpha\beta^{-{1\over 4}} |t|^{1\over
4}\eta_{\mu\nu}$ does not transform and remains large, but the
string metric, $g_{\mu\nu}^{(s)}={\alpha \over \sqrt
{\beta |t|}} \eta_{\mu\nu}$ shrinks to a point.

Another context in which our background can arise in string theory
is when there are some other compact dimensions.  Consider for
example the compactification of M theory on the flat space ${\bf
R}^8\times {\cal M}^2 \times {\cal S}^1$. That is,
we compactify one
of the Euclidean dimensions of the previously mentioned background
on a circle.  We consider the ${\cal S}^1$ factor as the M theory
circle, and interpret the theory as type IIA theory on ${\bf
R}^8\times {\cal M}^2$.  Since the size of the ${\cal S}^1$ is
independent of spacetime, the string coupling is constant.
Furthermore, since the metric on ${\bf R}^8\times {\cal M}^2$ is
flat (except perhaps at $t=0$) this is an exact solution of the
string equations of motion to all orders in $\alpha'$.  
As we explained above, the background breaks supersymmetry and,
therefore, we cannot argue that this is also an exact solution to
all orders in $g_s$.  Of particular interest are the winding
modes around the spatial circle in ${\cal M}^2$ near $t=0$. They
are reminiscent of the tachyonic winding modes which were
recently studied by Adams {\it et al.}\cite{adams}\null.
These modes might lead to an instability of our
system near the singularity and to a divergence of the loop
diagrams. One might be tempted to use T-duality to transform the
winding modes to momentum modes. The T-dual metric and string
coupling are $- (dt)^2 + {(\alpha')^2 \over B^4 t^2} (d \tilde w)^2$
and $\tilde g_s (t)=g_s{\sqrt {\alpha'} \over B^2 t}$. Both the
T-dual curvature and the T-dual string coupling are large for $t<
\sqrt {\alpha'}$ and, therefore, the T-dual picture is not
useful. The precise behavior of these modes and of other effects
near the singularity is a fascinating issue which we hope to
return to in a future publication.

\section{Cosmological Implications}

Our conjecture is that the Universe can undergo a transition from a
big crunch to a big bang by passing through a string theoretic regime
which connects the two phases.
It is standard in string compactifications
to have a singularity which is common to several
different classical spaces. 
The theory at the singularity is often less singular than one
might expect classically\cite{strominger}\null. The flop \cite{REFFLOP}
and the conifold \cite{REFCONI}  transitions
are particular examples of this general phenomenon. Even though
these singularities are spatial singularities, it is also likely that
dynamical singularities like our big crunch and big bang
singularities are similarly connected in string theory.
If so, what has been perceived as the beginning of time
may simply be a bridge to a pre-existing phase of the Universe.
The door is thereby opened to whole new classes of 
cosmological models, alternatives to the standard big bang and inflationary
models.

A particularly pertinent example is the recently proposed 
``ekpyrotic'' model of the universe \cite{ek1}\null.
According to this model, the 
universe began in a  non-singular, nearly vacuous
quasistatic state that lasted for an indefinite period.  The initial 
state can be described 
as a nearly BPS
(Bogolmon'yi-Prasad-Sommerfeld)
 configuration  of two orbifold boundary branes and 
 a 
(3+1)-dimensional brane in the bulk 
moving slowly along the intervening fifth dimension.
 The bulk brane is 
 attracted to   a boundary brane  by a force associated with 
a negative scalar potential. 
The radiation that fuels the hot big bang is 
generated in the collision between the 
branes. The BPS condition 
ensures that the universe is homogeneous and
spatially flat.
Ripples in the brane surface created by quantum fluctuations as the 
branes approach result in  a 
nearly scale-invariant spectrum of density perturbations after 
the collision.   In short, all of the cosmological problems of the 
standard hot big bang model are addressed.

For a bulk-brane/boundary-brane
collision,  the modulus that determines
the distance between the branes remains finite and gravity is only
a spectator. Consequently, this collision  
entails none of the
subtleties discussed in this paper. 
However, in order for the ekpyrotic model to be viable, there remains an 
important challenge.  In the long wavelength limit, the brane picture can be 
described by an effective 4d field theory with negative potential energy.  
Beginning from a static state, a negative potential energy causes the effective 
4d scale factor to shrink \cite{ek1}\null.  
In the braneworld picture, the Universe  continues to
shrink because the  boundary branes
are approaching one another\cite{ek1}\null.
It is essential that a mechanism exist that will 
reverse contraction to expansion after the bulk and boundary
branes collide, a point emphasized by many people
\cite{linde}\null.

In this paper,
we  have  focused on the reversal process and, particularly, on the possibility
of a  collision and rebound between the boundary branes.  
Our result suggests that the 
reversal to increasing $a$ might be accomplished by a second collision
between the boundary branes. The essence of our argument is that there exist 
variables that remain finite on each side of the 
bounce and that there is a natural way to  match across the bounce.
In Ref.~(\ref{lyth}), we discuss how
perturbations created during the contracting phase evolve into the 
expanding phase by identifying a set of perturbation variables that 
also remain finite at the bounce and naturally
match across the boundary.

A major modification of the ekpyrotic scenario suggests itself.
Perhaps the scenario can be accomplished with only the boundary branes
and no bulk brane.
Qualitatively, it is straightforward to show that,
if there is a negative, attractive potential drawing the two boundary    
branes towards one another which satisfies the conditions 
assumed before for the bulk brane-boundary brane potential, a
nearly scale-invariant spectrum of perturbations will be produced
that remains after the bounce, as discussed in Ref.~(\ref{lyth}).
We are currently examining this alternative scenario to determine 
if the quantitative requirements for the density perturbations can 
be met. If so, this would represent a significant simplification
relying on novel physical processes that occur when boundary branes collide.

\medskip

\noindent
{\bf Acknowledgements:} 

\noindent
We thank 
T. Banks, G. Horowitz, 
J. Maldacena, J. Polchinski and E. Witten  for useful discussions.
NT thanks N. Kaloper for bringing of Ref.~\ref{kal} to our attention.
This work was supported in part by
the Natural Sciences and Engineering Research Council of
Canada (JK),
  the  US Department of Energy grants
DE-FG02-91ER40671 (JK and PJS) and DE-AC02-76-03071 (BAO), 
DE-FG02-90ER40542 (NS),
and by PPARC-UK (NT).

\end{document}